\documentclass{journal}

\usepackage{natbib}
\usepackage{graphicx}
\usepackage{newtxtext}
\usepackage{newtxmath}
\usepackage{hyperref}
\hypersetup{colorlinks = true, urlcolor   = blue,
                               citecolor  = black,
                               linkcolor  = black}

\journal{Eur. J. Appl. Math.}

\title{The effect of local ventilation on a spatiotemporal model of airborne disease transmission in indoor spaces.}
\shorttitle{Effect of local ventilation on disease transmission}

\author{Alexander Pretty\aff{1},
        Ian M. Griffiths\aff{2},
        Zechariah Lau\aff{1,2}
   \and Katerina Kaouri\aff{1}}
\shortauthor{A. Pretty et al.}

\affiliation{
  \aff{1}School of Mathematics, Cardiff University, Cardiff, CF24 4AG, UK
  \aff{2}Mathematical Institute, University of Oxford, Oxford, OX2 6GG, UK}

\corremail{prettya@cardiff.ac.uk}

\begin{document}
\maketitle

\begin{keywords}
Epidemiology, Diffusion and convection, Mathematical modeling or simulation for problems pertaining to fluid mechanics, Reaction-diffusion equations
\end{keywords}

\begin{abstract}
  We incorporate local ventilation effects into a spatially dependent generalisation of the Wells--Riley model for airborne disease transmission. Aerosol production and removal through ventilation, biological deactivation, and gravitational settling as well as transport around a recirculating air-conditioning flow and turbulent mixing are modelled using an advection--diffusion--reaction equation. The local ventilation model, motivated by air purifiers, is compared with the global ventilation model for a weak purifier (CADR~=~140~m$^3$h$^{-1}$) and a strong purifier (CADR~=~1,000~m$^3$h$^{-1}$).

  We find that, as expected, increasing the distance of the infectious person from the purifier reduces the aerosol concentration. Moreover, the concentration is generally lowest when the infectious person is upstream of the purifier, located in regions where the airflow streamlines are directed into the purifier inlet. For these infectious source locations, the global ventilation model significantly overestimates the concentration throughout the room. For infectious sources outside of these regions, there is generally good agreement between the models, particularly for the weak purifier.

  We also studied, for fixed distance from the purifier, how the infection risk to a susceptible person varies as the infectious person changes location. The infection risk is greatest when the susceptible person is directly downstream of the infectious person. There is better agreement between local and global ventilation models for the weak purifier than the strong purifier.
\end{abstract}

\section{Introduction}

  The importance of ventilation in reducing the indoor transmission of infectious diseases was first highlighted by \citet{Nightingale}. When an infectious person breathes, talks, coughs, or sneezes, disease-carrying particles are emitted. In poorly ventilated spaces, small particles known as \emph{aerosols} can remain airborne for several hours, transmitting the disease to susceptible people when inhaled. During the COVID-19 pandemic, a key shift in understanding and mitigating transmission of the SARS-CoV-2 virus was the recognition of airborne transmission \citep{Morawska2020}, most likely responsible for superspreader outbreaks in a restaurant \citep{Lu2020,Ho2021b}, courtroom \citep{Vernez2021}, choir practice \citep{Miller2021}, and meat processing plant \citep{Gunther2020}.

  There are two main approaches to modelling airborne transmission: Wells--Riley models and Computational Fluid Dynamics (CFD). Wells--Riley models \citep{Riley78} assume a well-mixed-room (WMR), meaning aerosols are instantaneously transported throughout the room. Due to its high computational speed, this approach can be readily applied at the start of (and throughout) an epidemic. This was the case for COVID-19 \citep{Buonanno2020,Dai2020,Lelieveld2020}. However, the WMR assumption is not always appropriate and cannot provide any information on the spatial variation of the concentration around the room. One of the benefits of this simplified approach is how broadly applicable it is -- results do not depend on the specific configuration of people, furniture, and ventilation.

  CFD models simulate the detailed (usually turbulent) airflow in a room. The computational demand of these methods is high, so many CFD studies at the start of the COVID-19 pandemic focused on relatively short time frames (less than 5 minutes) \citep{Shafaghi2020,Vuorinen2020}. This is appropriate for heavier virus-laden droplets that quickly fall to the ground, but airborne transmission typically occurs over hours. Some CFD models have simulated aerosol evolution for up to an hour \citep{Shao2021}, but the high computational times means that CFD has a limited role in informing up-to-date decisions in a quickly developing epidemic. One key role of CFD models is in exploring a transmission event after-the-fact \citep{Ho2021b} to better understand the mechanisms influencing transmission.

  \citet{Lau2022} model the spatiotemporal evolution of aerosols in a room using an advection--diffusion--reaction (ADR) equation under the assumption of a recirculating airflow. Extending Wells--Riley models, the \citet{Lau2022} model accounts for the spatial variation of concentration and infection risk. Moreover, a simplified 2D airflow allows fast simulations so the model can be quickly deployed in a fast-changing epidemic. In \citet{Lau2022}, aerosol removal by ventilation is modelled as a global sink, as in the Wells--Riley type models. This assumption produces good agreement with real-life scenarios ventilated by inbuilt air-conditioning (AC) units \citep{Lau2022}.

  For rooms with poor or non-existent AC, air purifiers can increase overall aerosol removal. The effectiveness of air purifiers depends strongly on their location \citep{Burgmann2021,Narayanan2021}, an effect that cannot be captured by the WMR assumption or the global sink assumed in \citet{Lau2022}. Moreover, many CFD studies of air purifiers report prohibitively long computational times; for example \citet{Dbouk2021} report 7 days to simulate a 2.5-minute event in a domestic setting. While air purifiers are not technically classified as ventilation, since they do not provide fresh air from outdoors, the American Society of Heating, Refrigeration and Air-Conditioning (ASHRAE) incorporate air cleaning devices in their measure of \emph{equivalent clean airflow} when risk of disease transmission is high (e.g. during an epidemic) \citep{ASHRAE241}. In the present study, following ASHRAE, we conflate air purifiers with ventilation since we are solely interested in aerosol removal. However, we not that relying only on air purifiers with no outdoor air exchange can result in a build up of carbon dioxide, degrading the indoor air quality (IAQ).

  In this paper, we introduce spatially localised ventilation to the modelling framework of \citet{Lau2022}. While all ventilation systems (including doors, windows and AC units) have local effects, we are motivated by air purifier; we introduce a cylindrical device that draws air in through the top and expels clean air from the bottom. This device is placed in addition to an existing in-built (poorly performing) AC unit.

  This model allows quick investigation of transmission events over long periods of time. Lectures, meetings, school lessons, and certain time-limited social settings (e.g. escape rooms) usually last for about an hour. Time spent in restaurants, art galleries, networking events (e.g. conferences and workshops), and other social settings (e.g. pub quizzes) often last significantly longer. These scenarios are common transmission events for airborne diseases. We will focus on 1- to 4-hour events to explore the effectiveness of air purifiers in such settings.

  The paper is organised as follows. The modelling framework, incorporating a local ventilation system, is presented in \S \ref{Model}. In \S \ref{AverageC}, we compare the average aerosol concentration in the room predicted by the local and global ventilation models. The spatial distribution of aerosols is then compared in \S \ref{Spatial}. In \S \ref{Susceptible} we consider the infection risk at specific locations arising from the two ventilation models. Conclusions and suggestions for future work are provided in \S \ref{Conclusions}.

\section{Modelling framework} \label{Model}

  \subsection{Advection--diffusion--reaction (ADR) equation}

\begin{figure}
      \includegraphics[width=0.49\textwidth]{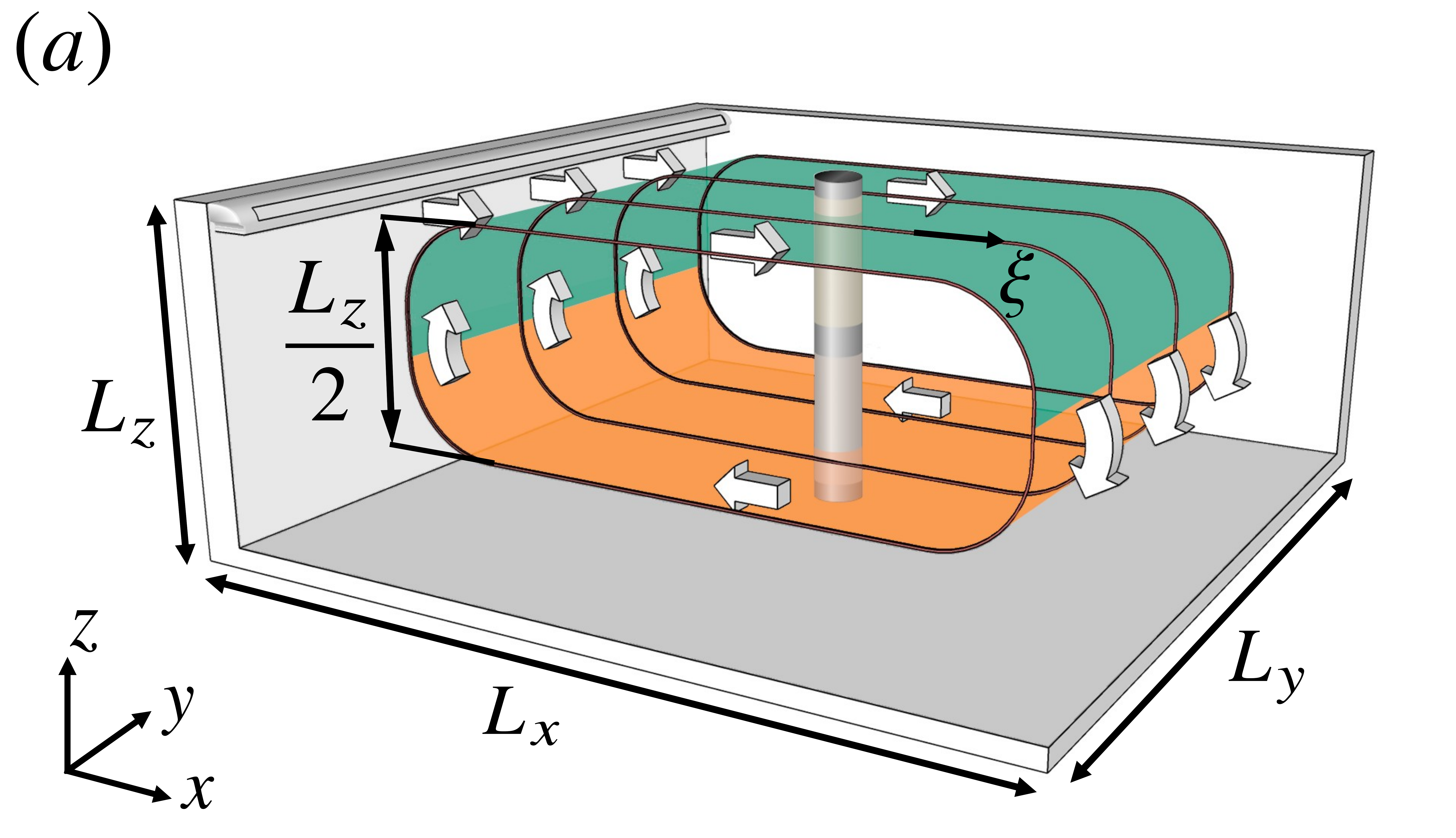}%{Fig1a.eps}
      \includegraphics[width=0.49\textwidth]{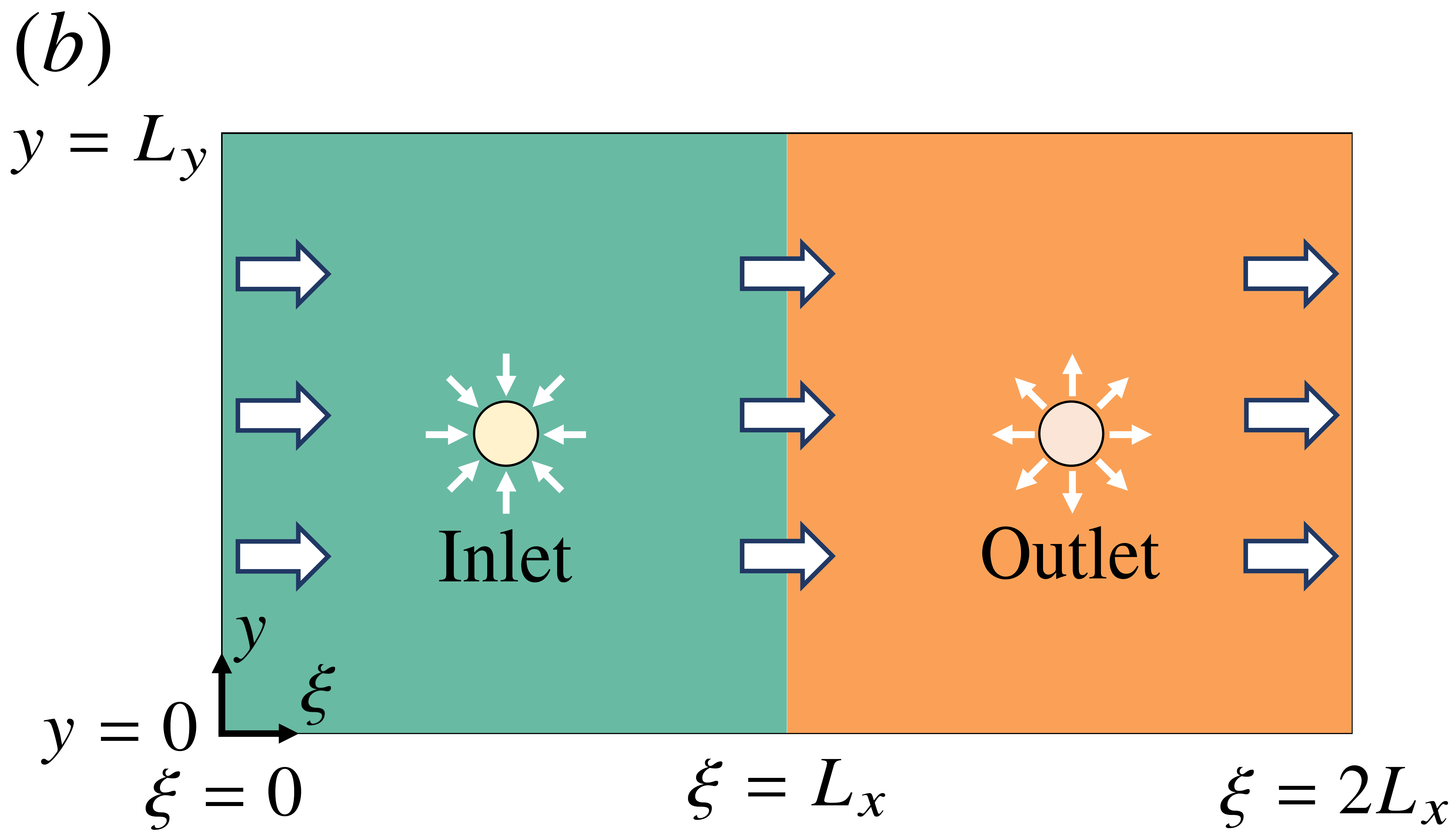}%{Fig1b.eps}

      \caption{A 3D room with dimensions $L_x$, $L_y$, $L_z$ is shown in (\textit{a}). A cylindrical local ventilation system crosses the two recirculating layers and the arclength coordinate $\xi$ follows the recirculating loop. The computational domain $(\xi,y)$ is shown in (\textit{b}).} \label{Fig1}
    \end{figure}

    Consider a three-dimensional (3D) room of volume $V$, with dimensions $L_x$, $L_y$, $L_z$, as depicted in Figure~\ref{Fig1}(\textit{a}). Following \citet{Lau2022}, we assume a recirculating flow produced by a single AC vent along the top corner and introduce the arclength coordinate $\xi$, which follows the recirculating loop. The distance between the recirculation layers is $L_z/2$ \citep{Hooff13} and the total arclength is $2L_x$. A cylindrical local ventilation system with radius $r$ is introduced, which extends over both recirculating layers: Air is drawn into an inlet in the upper layer and expelled from an outlet in the lower layer. We will refer to this device as a \emph{purifier}.

    Figure~\ref{Fig1}(\textit{b}) shows the computational domain $(\xi,y$). The left/right halves correspond to the upper/lower layers and the purifier inlet and outlet appear as circles removed from the domain, with boundaries
    \begin{subequations} \label{eq:PurBound}
      \begin{gather}
        \partial_{\mathrm{in}} = \{(\xi,y) : |(\xi,y) - (x_p,y_p)| = r\}, \\
        \partial_{\mathrm{out}} = \{ (\xi,y) : |(\xi,y) - (2L_x-x_p,y_p)| = r\},
      \end{gather}
    \end{subequations}
    where $(x_p,y_p$) denotes the location of the centre of the purifier in the $(x,y)$-plane, and we will assume throughout this work that the purifier is in the centre of the room. The purifier design is motivated by real-life devices \citep{Dbouk2021}. However, in this quasi-3D model the device extends the entire height of the room, which is significantly taller than real-life purifiers. This simplification allows for a direct comparison with the model of \citet{Lau2022} while still offering useful and practical insights into the local effects of air purifiers.

    Consider a single infectious individual at $\boldsymbol{x}_0 = (x_0,y_0)$ and talking continuously. Talking produces around 10 times as many aerosols as breathing \citep{Asadi19}: a reasonable worst-case scenario for an asymptomatic person. The infectious source will be placed in the top recirculating loop ($\xi = x_0$), meaning the infectious person is standing (e.g. a presenting teacher or lecturer). We assume, as in \citet{Lau2022}, that the concentration of aerosols, $\mathcal{C}(\xi,y,t)$, is governed by the ADR equation
      \begin{equation} \label{eq:ADR}
        \frac{\partial \mathcal{C}}{\partial t} + \nabla \cdot ( \boldsymbol{v} \mathcal{C} ) - \nabla \cdot ( K \nabla \mathcal{C}) = R \delta(\xi - x_0) \delta(y - y_0) - (\lambda + \beta + \sigma) \mathcal{C},
      \end{equation}
    where $\boldsymbol{v} = (u,v)$ is a vector field describing the airflow around the recirculating loop; $K$ is the eddy diffusion coefficient; $R$ is the (constant) aerosol production rate; and $\lambda$, $\beta$, $\sigma$ are global removal rates due to ventilation, biological deactivation, and gravitational settling, respectively. Parameter values are given in Table~\ref{Parameters}; more details are provided in \citet{Lau2022}.

    Following \citet{Lau2022}, we model the inbuilt ventilation with the global removal term $\lambda$ and set $\lambda = 2 \times 10^{-4}$~s$^{-1}$; the `poor ventilation' scenario of \citet{Lau2022} \citep[from classroom data,][]{Guo2008}. This reflects a broken or poorly maintained AC system that moves air around but is ineffective at removing aerosols \citep{Lu2020} with an air exchange rate of 0.7 air hanges per hour (ACH). (ACH will be given to 1 decimal place throughout this paper.)

    \begin{table}
      \caption{Parameters and their values.} \label{Parameters}
      \begin{center}
        \begin{tabular}{r c l l}
          \hline
          Parameter                    & Symbol    & Value                         & Source                           \\
          \hline
          Room length                  & $L_x$     & 8~m                           & \citet{Lau2022}                  \\
          Room width                   & $L_y$     & 8~m                           & \citet{Lau2022}                  \\
          Room height                  & $L_z$     & 3~m                           & \citet{Lau2022}                  \\
          Room volume                  & $V$       & 192~m$^3$                     & $L_x \times L_y \times L_z$      \\[3pt]
          AC airflow speed             & $u_0$     & 0.15~ms$^{-1}$                & \citep{ASHRAE55}                 \\[3pt]
          Aerosol emission rate (talking) & $R$    & 5~aerosols/s                  & \citep{Lau2022}                  \\[3pt]
          Virus deactivation rate      & $\beta$   & $1.7\times10^{-4}$~s$^{-1}$   & \citep{Doremalen2020}            \\
          Gravitational settling rate  & $\sigma$  & $1.1\times10^{-4}$~s$^{-1}$   & \citep{DeOliveria2021}           \\[3pt]
          Air exchange rate            & $\lambda$ & 0.7 ACH: $2\times10^{-4}$     & \citep{Guo2008}                  \\
          (s$^{-1}$)                   &           & 1.4 ACH: $4.0\times10^{-4}$   & see Table~\ref{PurifierParam}    \\
                                       &           & 6 ACH: $1.7\times 10^{-3}$    & see Table~\ref{PurifierParam}    \\[3pt]
          Eddy diffusion coefficient   & $K$       & 0.7 ACH: $5.3\times10^{-3}$   & \citep{Foat2020},\eqref{eq:Ktot} \\
          (m$^2$s$^{-1}$)              &           & 1.4 ACH: $1.0\times10^{-2}$   & \citep{Foat2020},\eqref{eq:Ktot} \\
                                       &           & 6 ACH: $4.5\times10^{-2}$     & \citep{Foat2020},\eqref{eq:Ktot} \\[3pt]
          Breathing rate             & $\rho$ & $1.3\times10^{-4}$~m$^{3}$s$^{-1}$ & \citep{Hallett2020}              \\[3pt]
          Infectivity constant         & $I$       & 0.0069                        & \citep{Lau2022}                  \\[3pt]
          Location of purifier centre  & $(x_p,y_p)$ & (4,4)~m                     & Room centre: $(L_x/2, L_y/2)$    \\
          Radius of purifier           & $r$       & 0.1~m                         &                                  \\
          \hline
        \end{tabular}
      \end{center}
    \end{table}

    Aerosol production begins at $t = 0$, so we set the initial condition $\mathcal{C}(\xi,y,0) = 0$. Periodic conditions across the left and right boundaries ($\xi = 0, 2L_x$) complete the recirculating loop,
    \begin{subequations}
      \begin{gather}
        \mathcal{C}(0,y,t) = \mathcal{C}(2L_x,y,t), \\
        \frac{\partial \mathcal{C}}{\partial \xi} (0,y,t) = \frac{\partial \mathcal{C}}{\partial \xi} (2L_x,y,t),
      \end{gather}
    and there is no flux through the walls at $y = 0, L_y$,
      \begin{gather}
        \frac{\partial \mathcal{C}}{\partial y}(\xi,0,t) = \frac{\partial \mathcal{C}}{\partial y}(\xi,L_y,t) = 0.
      \end{gather}
    \end{subequations}
    Aerosols are carried out of the purifier inlet by advection, so we set zero diffusive flux,
    \begin{subequations}
      \begin{gather}
        \boldsymbol{\hat{n}} \cdot (K \nabla \mathcal{C}) = 0 \; \; \mathrm{on} \; \; \partial_{\mathrm{in}}.
      \end{gather}
    No aerosols enter the domain through the purifier outlet, so we set zero total (diffusive and advective) flux,
      \begin{gather}
        \boldsymbol{\hat{n}} \cdot ( \boldsymbol{v} \mathcal{C} - K \nabla \mathcal{C}) = 0 \; \; \mathrm{on} \; \; \partial_{\mathrm{out}}.
      \end{gather}
    \end{subequations}
    Here, $\boldsymbol{\hat{n}}$ denotes the unit vector normal to each boundary, directed out of the domain.

    Extending the formula for the diffusion coefficient, $K$, from \citet{Foat2020}, we assume that $K$ is related to the total air exchange rate as follows,
    \begin{equation} \label{eq:Ktot}
      K = (\lambda + \lambda_p) \sqrt[3]{\frac{V^2}{2}},
    \end{equation}
    where $\lambda_p$ is the air exchange rate of the purifier (discussed below). The value of $K$ is therefore the same for equivalent global and local ventilation levels. This makes direct comparison straightforward, but does not account for different vent sizes generating different-sized turbulent eddies.

    Following \citet{Lau2022}, we assume that the majority of aerosols remain within the recirculating loop and are well-mixed over this height. Hence, the concentration in aerosols/m$^3$ is given by
    \begin{equation}
      C(x,y,t) = \frac{\mathcal{C}(\xi = x,y,t) + \mathcal{C}(\xi = 2L_x - x,y,t)}{L_z/2}.
    \end{equation}

  \subsection{Infection risk}

    The risk of infection to a susceptible person at any $(x,y)$ is then calculated using
    \begin{equation} \label{eq:InfRisk}
      P(x,y,t) = 1 - \mathrm{exp} \left[ - I \int_0^t \rho C(x,y,\tau) \, \mathrm{d} \tau \right],
    \end{equation}
    where $I$ is the infectivity constant of the virus and $\rho$ is the breathing rate (see Table~\ref{Parameters}).

  \subsection{Airflow}

    In \citet{Lau2022}, aerosols are advected around the recirculating loop at constant speed $u_0$ (Table~\ref{Parameters}). Here, we assume that the recirculating loop remains coherent in the presence of a purifier, but we do not assume constant velocity. This leads to a modified $\boldsymbol{v} = (u,v)$ such that air enters and leaves the purifier with a specified constant speed $v_p$:
    \begin{subequations} \label{eq:FlowInOut}
      \begin{gather}
        \boldsymbol{v} \cdot \boldsymbol{\hat{n}} = v_p \; \; \mathrm{on} \; \; \partial_{\mathrm{in}}, \\
        \boldsymbol{v} \cdot \boldsymbol{\hat{n}} = - v_p \; \; \mathrm{on} \; \; \partial_{\mathrm{out}}.
      \end{gather}
    \end{subequations}
    We also require that $\boldsymbol{v}$ satisfies the periodic boundary condition,
    \begin{equation} \label{eq:FlowPBC}
      \boldsymbol{v}(0,y) = \boldsymbol{v}(2 L_x,y).
    \end{equation}

    The velocity vector field that satisfies \eqref{eq:FlowInOut} and \eqref{eq:FlowPBC} is determined using the Shear Stress Transport (SST) turbulent flow solver \citep{Menter1994} in COMSOL Multiphysics$^{\circledR}$ \citep[see][]{comsol}. A turbulent flow solver is used since $Re > 10,000$ and hence numerical solutions to the Navier--Stokes equations were found to be sensitive to the mesh size. The resulting 2D flow in the $(\xi,y)$-plane does not incorporate the diffusive properties of a true 3D turbulent flow; the spreading of aerosols by small-scale turbulent eddies is accounted for by the eddy diffusion coefficient $K$, discussed above.

    \begin{table}
      \begin{center}
        \caption{Parameters for two purifiers, with $\lambda = 2\times10^{-4}$~s$^{-1}$ (0.7 ACH).} \label{PurifierParam}
        \begin{tabular}{r c l l l}
          \hline
          Parameter                  & Symbol          & Weak purifier                     & Strong purifier             & Source              \\
          \hline
          CADR                       &                 & 140~m$^3$h$^{-1}$                 & 1,000~m$^3$h$^{-1}$         & *                    \\
          Flow-rate                  & $Q$             & 0.039~m$^3$s$^{-1}$               & 0.28~m$^3$s$^{-1}$          & CADR$/$3,600        \\[4pt]
          Air velocity into purifier & $v_p$           & 0.04~ms$^{-1}$                    & 0.3~ms$^{-1}$               & (\ref{eq:PurQ})     \\
          Purifier air exchange rate & $\lambda_p$     & $2.0\times10^{-4}$~s$^{-1}$       & $1.5\times10^{-3}$~s$^{-1}$ & $Q/V$               \\
          Total air exchange rate & $\lambda_{\mathrm{tot}}$ & $4.0\times10^{-4}$~s$^{-1}$ & $1.7\times10^{-3}$~s$^{-1}$ & $\lambda+\lambda_p$ \\[4pt]
          Equivalent global ACH      &                 & 1.4 ACH                           & 6.0 ACH           & 3,600$\lambda_{\mathrm{tot}}$ \\
          \hline
        \end{tabular}

        * Weak purifier: \citet{Dbouk2021}, Strong purifier: \citet{Kahler2020a}.
      \end{center}
    \end{table}

    Imposing no-slip and no-penetration conditions at the walls,
    \begin{equation}
      \boldsymbol{v}(\xi,0) = \boldsymbol{v}(\xi,L_y) = (0,0),
    \end{equation}
    we run the SST solver until a steady state is reached. This steady velocity, $\boldsymbol{v}$, is then used in the ADR equation \eqref{eq:ADR}. The results are compared against those of \citet{Lau2022} for: (i) no purifier, and (ii) a switched-off purifier ($v_p = 0$). There is good agreement in both cases provided
    \begin{equation}
      \max_y u (\xi = 0, y) = u_0,
    \end{equation}
    which is imposed by setting a suitable pressure gradient over the periodic boundaries. Several turbulent models were compared and all resulted in a similar concentration distribution $\mathcal{C}$.

    Two purifier settings are considered based on the clean air delivery rate (CADR) of purifiers used in experimental and computational studies. We define a \textit{weak purifier} with a CADR of 140~m$^{3}$h$^{-1}$, representative of devices for small spaces such as domestic rooms \citep{Dbouk2021} and individual offices; and a \textit{strong purifier} with a CADR of 1,000~m$^{3}$h$^{-1}$, representative of devices for larger spaces such as classrooms \citep{Kahler2020a} and open-plan offices.

    Let $Q$ denote the flow-rate through the device in m$^3$s$^{-1}$. The CADR (stated in m$^3$h$^{-1}$ by convention) is given by $\eta Q$ where $\eta$ is the filter efficacy. We assume that 100\% of the aerosols that enter the purifier are trapped by the filter, so $\eta = 1$ and the CADR and flow-rate $Q$ are equivalent. For a cylindrical purifier (circumference $2 \pi r$) with an inlet half the height of the room ($L_z/2$),
    \begin{equation} \label{eq:PurQ}
      Q = \pi r L_z v_p.
    \end{equation}
    Using $v_p = Q/\pi r L_z$ as the air speed at the inlet and outlet boundaries (see Table~\ref{PurifierParam}), we determine the velocity field $\boldsymbol{v}$ for each purifier.

    Each purifier is compared with an equivalent increase in the global removal term $\lambda$. The air exchange rate associated with each purifier is given by $\lambda_p = Q/V$, which we add to the global removal rate of the AC unit (0.7 ACH) to determine a total air exchange rate, $\lambda_{\mathrm{tot}} = \lambda + \lambda_p$ (see Table~\ref{PurifierParam}). The weak purifier doubles the total air exchange to 1.4 ACH, less than half the recommended ventilation for classrooms \citep[3 ACH for 30 occupants:][]{ASHRAE62}. The total air exchange rate for the strong purifier is 6 ACH, exceeding this recommendation and also sufficient to meet the guidelines for times of heightened infection risk, provided the number of occupants is halved \citep{ASHRAE241}. Hereon in, we will refer to the global ventilation model by the ACH and local ventilation model by the purifier strength.

    The global ventilation cases will include a switched-off purifier ($v_p = 0$) since removing the purifier entirely results in a different computational domain to the purifier cases. This has little impact on the concentration around the room. The control case of 0.7 ACH (only the AC unit) will be referred to as the \emph{no purifier} case.

\begin{figure}
      \includegraphics[width=0.49\textwidth]{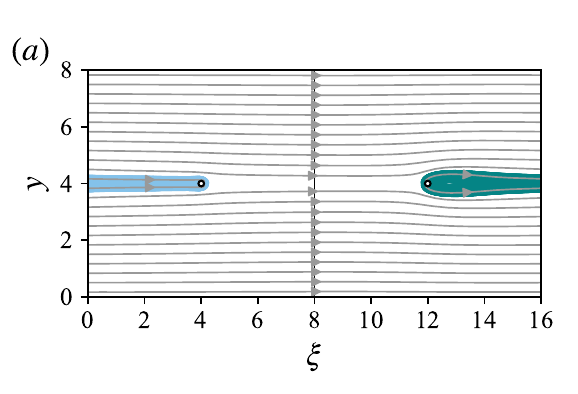}%{Fig2a.eps}
      \includegraphics[width=0.49\textwidth]{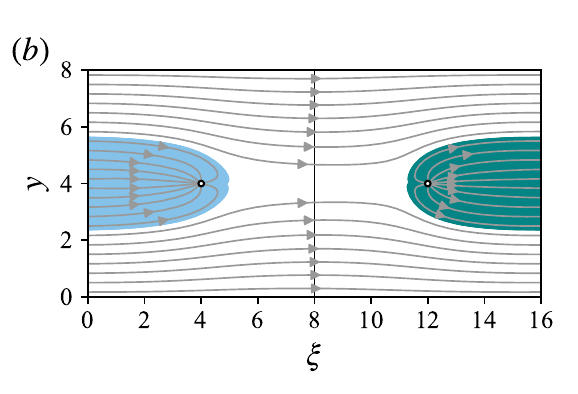}%{Fig2b.eps}
      \vspace{-1ex}

      \caption{The airflow streamlines in the $(\xi,y)$-plane are shown for (\textit{a}) the weak purifier ($v_p = 0.04$) and (\textit{b}) the strong purifier ($v_p = 0.3$). The shaded regions indicate streamlines that pass through the purifier inlet (left) and the purifier outlet (right).}
      \label{Fig2}
    \end{figure}

    The streamlines of the airflow $\boldsymbol{v}$ are shown in Figure~\ref{Fig2} for both purifiers. The flow is broadly unidirectional for the weak purifier (Figure~\ref{Fig2}\textit{a}). Regions in which the streamlines are directed into or out of the purifier are shaded and take up a greater area for the strong purifier (Figure~\ref{Fig2}\textit{b}). In both cases, these regions meet at the periodic boundary.

  \subsection{Computational speed}

    Simulations were performed on a Lenovo IdeaPad Flex 5 laptop, with a 1.3 GHz 4-core Intel Core i7-1065G7 processor and 8 GB of RAM. For each airflow simulation (no purifier, weak purifier, strong purifier) to reach a steady state, computation takes approximately 10 minutes. For the ADR equation \eqref{eq:ADR}, an event of 4 hours takes around 5 minutes to run, including calculation of the infection risk \eqref{eq:InfRisk}. At these computational speeds, advice and guidance can be quickly updated with new information during a fast-changing epidemic.

\section{Average aerosol concentration} \label{AverageC}

  \begin{figure}
    \includegraphics[width=0.49\textwidth]{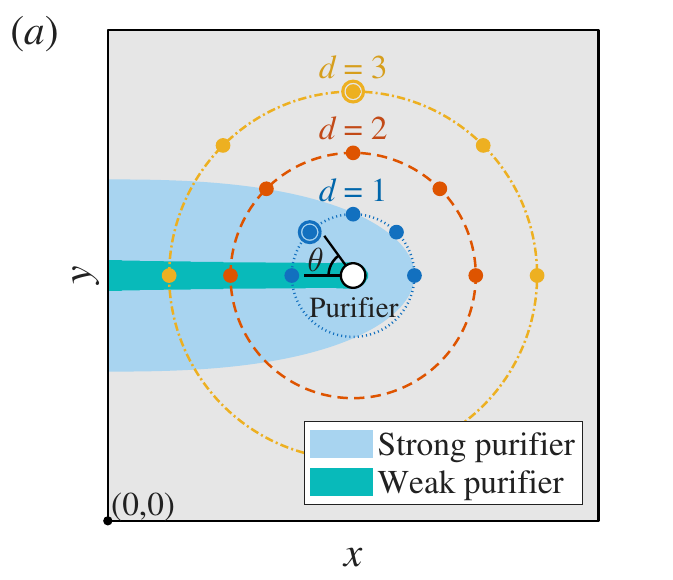}%{Fig3a.eps}
    \includegraphics[width=0.49\textwidth]{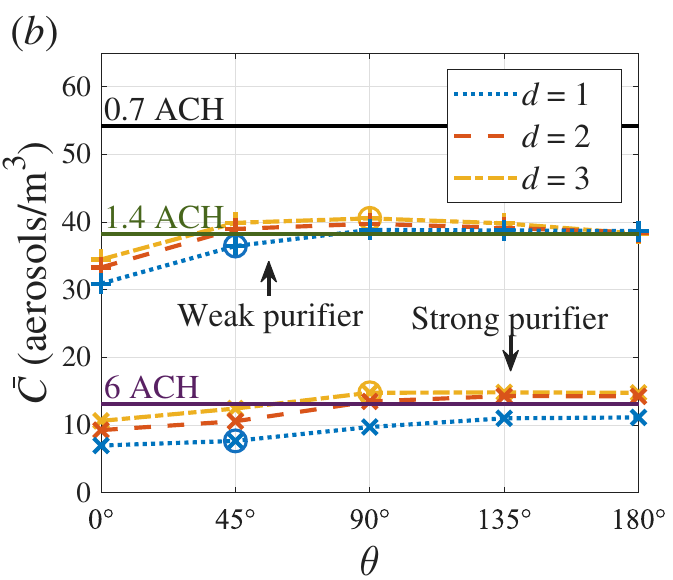}%{Fig3b.eps}
    \caption{In (\textit{a}), the infectious source locations, $\boldsymbol{x}_0$, are depicted by filled circles and the regions where all streamlines are directed into the purifier inlet (Figure~\ref{Fig2}) are shaded for each purifier. In (\textit{b}), the average aerosol concentration, $\bar{C}$, after 4 hours is shown for the weak ($+$) and the strong ($\times$) purifiers for these $\boldsymbol{x}_0$. The global ventilation cases are also shown in (\textit{b}), depicted as horizontal lines (labelled with the ACH). The cases explored further in Figures \ref{Fig4} and \ref{Fig5} are indicated with circles.} \label{Fig3}
  \end{figure}

  We define the average aerosol concentration in the room as
  \begin{equation} \label{eq:Cbar}
    \bar{C}(t) = \frac{1}{V} \iint_{\Omega} \mathcal{C}(\xi,y,t) \, \mathrm{d} \xi \, \mathrm{d} y,
  \end{equation}
  where $\Omega$ denotes the computational domain depicted in Figure~\ref{Fig1}(\textit{b}). Taking appropriate integrals of \eqref{eq:ADR} and applying the divergence theorem gives
  \begin{equation} \label{eq:AvgCEq}
    \frac{\partial \bar{C}}{\partial t} = \frac{R}{V} - (\lambda + \beta + \sigma) \bar{C} - \frac{1}{V} \oint_{\partial_{\mathrm{in}}} v_p \mathcal{C} \, \mathrm{d} l,
  \end{equation}
  where $l$ denotes a line element on $\partial_{\mathrm{in}}$. When $v_p \neq 0$, the boundary integral (describing aerosol removal by the purifier) depends on the values of $\mathcal{C}$ on the purifier inlet boundary $\partial_{\mathrm{in}}$ \eqref{eq:PurBound}, which depend on the distribution of aerosols around the room. Hence, $\bar{C}$ implicitly depends on the location of the infectious source $\boldsymbol{x}_0$ and strength of the purifier. When $v_p = 0$ there is no local ventilation \citep[as in][]{Lau2022}, the boundary integral vanishes, and \eqref{eq:AvgCEq} is equivalent to the Wells--Riley model used in \citet{Miller2021}. In this case the solution to \eqref{eq:AvgCEq} is given by
  \begin{equation} \label{eq:SSavg}
    \bar{C}(t) = C^* \left[1 - \mathrm{e}^{-(\lambda+\beta+\sigma)t} \right], \; \; \mathrm{where} \; \; C^* = \frac{R}{(\lambda+\beta+\sigma)V}.
  \end{equation}
  Hence, $\bar{C}$ does not depend on $\boldsymbol{x}_0$ or $\boldsymbol{v}$ for the global ventilation cases, and $\bar{C} \rightarrow C^*$ as $t \rightarrow \infty$.

  We express the location of the infectious source relative to the purifier location as
  \begin{equation} \label{eq:InfLoc}
    \boldsymbol{x}_0 = (x_0,y_0) = (x_p - d \, \mathrm{cos} \, \theta, y_p + d \, \mathrm{sin} \, \theta),
  \end{equation}
  where $d$ is the distance from the purifier and $\theta$ is the angle (in degrees) from the line $y = L_y/2$. The problem is symmetric in this line so we consider only $\theta \in [0^{\circ},180^{\circ}]$. In Figure~\ref{Fig3}(\textit{a}), we show all choices of $(d,\theta)$ considered here.

  To compare the long-time behaviour of the different ventilation models, $\bar{C}$ is plotted in Figure~\ref{Fig3}(\textit{b}) after an event of 4 hours. This duration is representative of certain social settings (e.g. restaurants), activities within all-day events (such as conferences, workshops and networking events), and time spent working in open plan offices. For all the cases considered here, $\mathcal{C}$ reaches a steady state within this time.

  For the global ventilation models, $\bar{C} = C^*$ \eqref{eq:SSavg}, which we evaluate for 0.7 ACH, 1.4 ACH, and 6 ACH; plotted as horizontal lines in Figure~\ref{Fig3}(\textit{b}). For most $(d,\theta)$ choices, there is good agreement between each purifier and the equivalent global ventilation. For the weak purifier, the only significant deviation is when $\theta = 0$ (for all values of $d$). For the strong purifier, the discrepancy is significant for $d = 1$ (for all values of $\theta$) and for small $\theta$ when $d = 2,3$. Where the results differ the most, the global ventilation overestimates $\bar{C}$ compared to the local purifier model. When $d = 2, 3$, the global ventilation model underestimates $\bar{C}$ for some $\theta$, but the discrepancy is relatively small.

  In Figure~\ref{Fig3}(\textit{b}), we show that $\bar{C}$ increases with $d$ for both purifiers; purifiers are less effective when the infectious person is further away (as expected). Also, both purifiers are more effective (with $\bar{C}$ decreasing) when the infectious person is located in the shaded region where all streamlines enter the purifier inlet. In these cases, the global ventilation overestimates the concentration. When $\boldsymbol{x}_0$ is outside of these regions, there is less variation in $\bar{C}$ and closer agreement between the ventilation models.

  \begin{figure}
	  \includegraphics[width=0.49\textwidth]{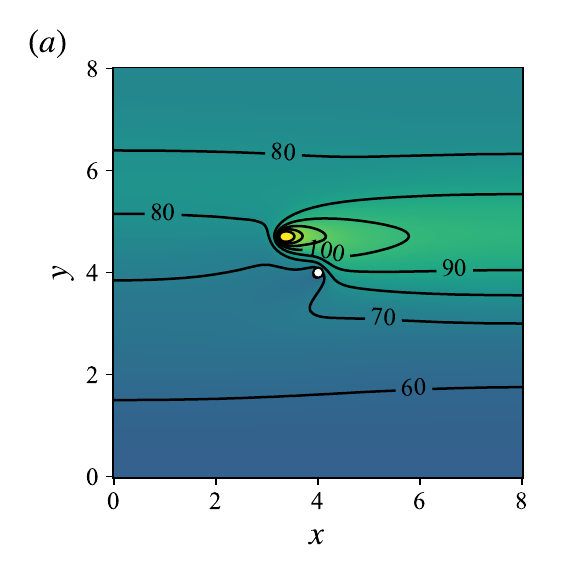}%{Fig4a.eps}
	  \includegraphics[width=0.49\textwidth]{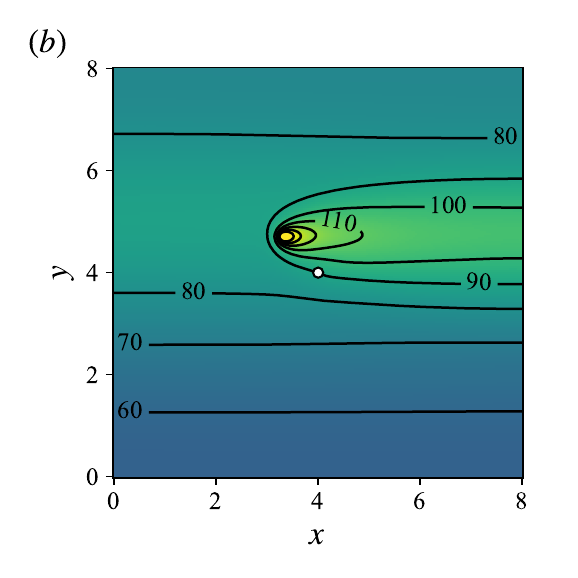}%{Fig4b.eps}
	  \vspace{-2ex}

	  \includegraphics[width=0.49\textwidth]{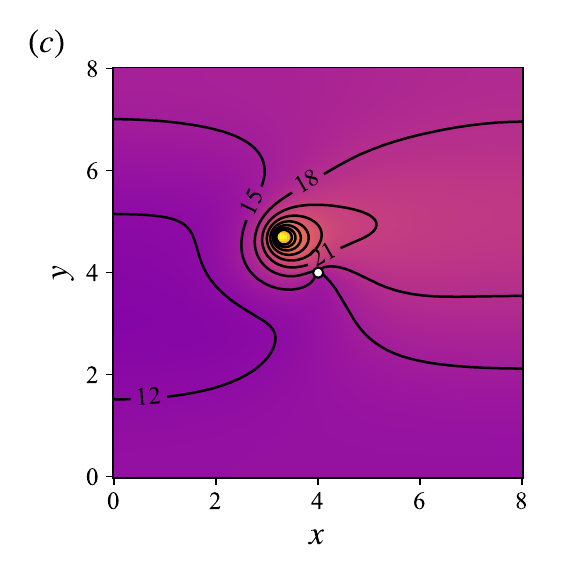}%{Fig4c.eps}
	  \includegraphics[width=0.49\textwidth]{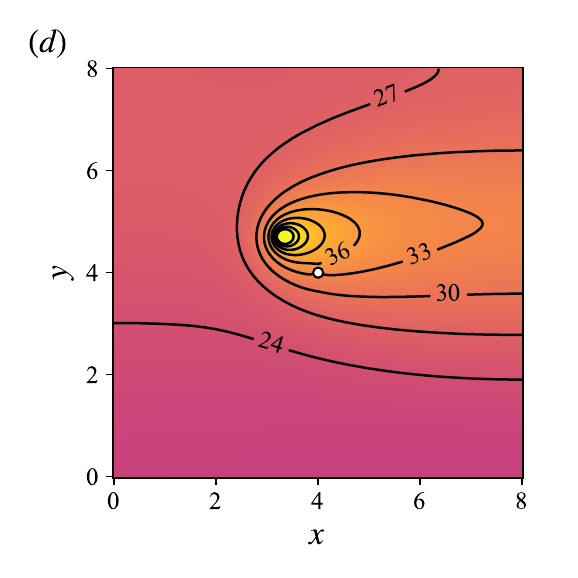}%{Fig4d.eps}
    \vspace{-2ex}

    \caption{Contour plots of the aerosol concentration $C$ after 4 hours for an infectious source at $\boldsymbol{x}_0$ \eqref{eq:InfLoc} with $(d,\theta) = (1,45^{\circ})$ (see Figure~\ref{Fig3}a). The concentration is shown for (\textit{a}) the weak purifier ($C_p$) and (\textit{b}) the equivalent global ventilation of 1.4~ACH ($C_g$), both with contours at intervals of 10~aerosols/m$^3$. The concentration is shown for (\textit{c}) the strong purifier ($C_p$) and (\textit{d}) the equivalent global ventilation of 6~ACH ($C_g$), both with contours at intervals of 3~aerosols/m$^3$.} \label{Fig4}
  \end{figure}

\section{Spatial distribution of aerosols} \label{Spatial}

\begin{figure}
	  \includegraphics[width=0.49\textwidth]{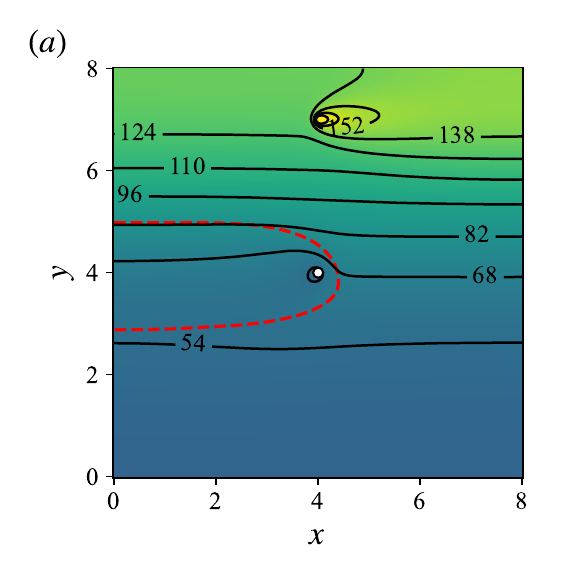}%{Fig5a.eps}
	  \includegraphics[width=0.49\textwidth]{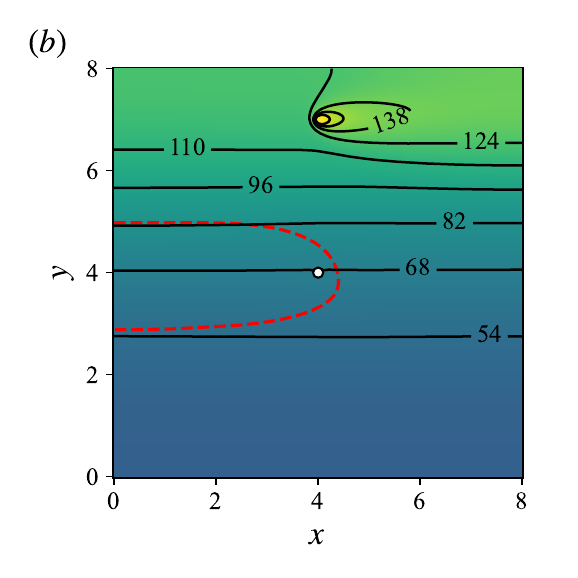}%{Fig5b.eps}
	  \vspace{-2ex}

	  \includegraphics[width=0.49\textwidth]{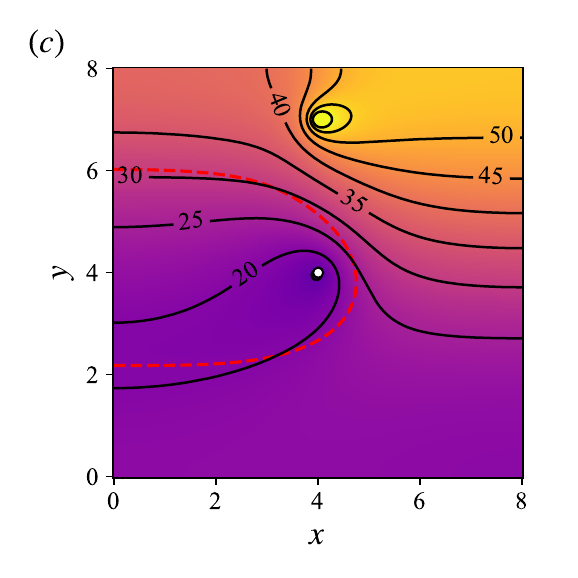}%{Fig5c.eps}
	  \includegraphics[width=0.49\textwidth]{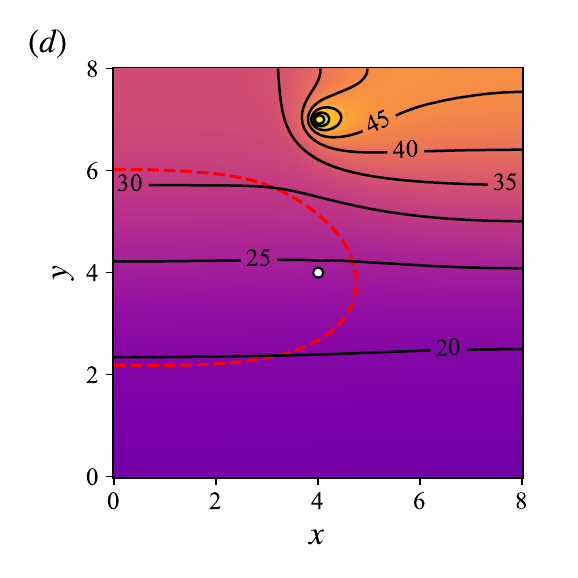}%{Fig5d.eps}
    \vspace{-2ex}

    \caption{Contour plots of the aerosol concentration $C$ after 4 hours for an infectious source at $\boldsymbol{x}_0$ \eqref{eq:InfLoc} with $(d,\theta) = (3,90^{\circ})$ (see Figure~\ref{Fig3}a). The concentration is shown for (\textit{a}) the weak purifier ($C_p$) and (\textit{b}) the equivalent global ventilation of 1.4 ACH ($C_g$), both with contours at intervals of 14~aerosols/m$^3$. The concentration is shown for (\textit{c}) the strong purifier ($C_p$) and (\textit{d}) the equivalent global ventilation of 6 ACH ($C_g$), both with contours at intervals of 5~aerosols/m$^3$. Additional dashed lines show where $C_p = C_g$.} \label{Fig5}
  \end{figure}

  The spatiotemporal model also provides information on the spatial distribution of aerosols around the room. The local and global ventilation models were compared for the infectious source locations depicted in Figure~\ref{Fig3}(\textit{a}). Contour plots showing the spatial variation of $C$ after 4 hours (as in Figure~\ref{Fig3}) are given in Figures~\ref{Fig4}~and~\ref{Fig5} for some representative examples. For this discussion, we will denote the concentration for the local ventilation (purifier) by $C_p$ and for the equivalent global ventilation by $C_g$.

  The concentration around the room is shown in Figure~\ref{Fig4} when $(d,\theta) = (1,45^{\circ})$ for both purifiers and the equivalent global ventilations. In Figures~\ref{Fig4}(\textit{a,b}), $C_p$ and $C_g$ are shown (respectively) for the weak purifier. There is a similar concentration throughout the room for both ventilation models but $C_p < C_g$ everywhere. The largest discrepancies are: below and to the left of the purifier; and to the right of the infectious source. Similar features were observed for all cases in which the infectious source is located near to (but outside of) the region where streamlines are directed into the purifier inlet (Figure~\ref{Fig3}\textit{a}).

  Figures~\ref{Fig4}(\textit{c,d}) show $C_p, C_g$ (respectively) for the strong purifier. In this case, the global ventilation model significantly overestimates the concentration compared to the local ventilation model. For instance, $C_g > 21$~aerosols/m$^3$ everywhere in Figure~\ref{Fig4}(\textit{d}) whereas $C_p < 21$~aerosols/m$^3$ in the majority of the room in Figure~\ref{Fig4}(\textit{c}). These observations are indicative of all cases where the infectious source lies inside the region where streamlines are directed into the purifier inlet (see Figure~\ref{Fig3}\textit{a}).

  In Figure~\ref{Fig5}, we consider the distribution of aerosols when $(d,\theta) = (3,90^{\circ})$. These are representative examples for the cases where the infectious source is far away from the region where the streamlines are directed into the purifier inlet.

  Comparing $C_p$ and $C_g$ for the weak purifier (Figures~\ref{Fig5}\textit{a,b}), the concentration is similar throughout the room, except for the region $y > 5$, where $C_p$ is approximately one contour level (14~aerosols/m$^3$) larger than $C_g$. There are regions in the room where $C_p < C_g$ (near the purifier) and where $C_p > C_g$ (near the infectious source), with the dashed red line (the same in both plots) showing the boundary between these regions, i.e.~where $C_p = C_g$ (no such lines appear in Figure~\ref{Fig4} since $C_p < C_g$ everywhere in those cases).

  For the strong purifier example shown in Figures~\ref{Fig5}(\textit{c,d}), the spatial distribution of aerosols differs more significantly between the ventilation models. The region where $C_p < C_g$ is larger in this case, but where $C_p > C_g$ the concentration difference ranges from around 1 contour level (5~aerosols/m$^3$), in the top left and bottom right of the room, to around 2 contour levels, in the top right (near to the infectious source). This suggests that the local ventilation model better captures the aerosol distribution of the weak purifier than the strong purifier.

\section{Infection risk to susceptible people} \label{Susceptible}

  Our spatially varying model allows us to determine the infection risk to susceptible people at specific locations, \eqref{eq:InfRisk}. We express the location of a susceptible person $\boldsymbol{x}_s$ as
  \begin{equation} \label{eq:SusLoc}
    \boldsymbol{x}_s = (x_s,y_s) = (x_p - d_s \, \mathrm{cos} (\theta + \phi), y_p + d_s \, \mathrm{sin}(\theta + \phi)),
  \end{equation}
  where $\phi$ is the angle between $\boldsymbol{x}_0$ and $\boldsymbol{x}_s$, and $d_s$ is the distance from the purifier.

  \begin{figure}
    \includegraphics[width=0.49\textwidth]{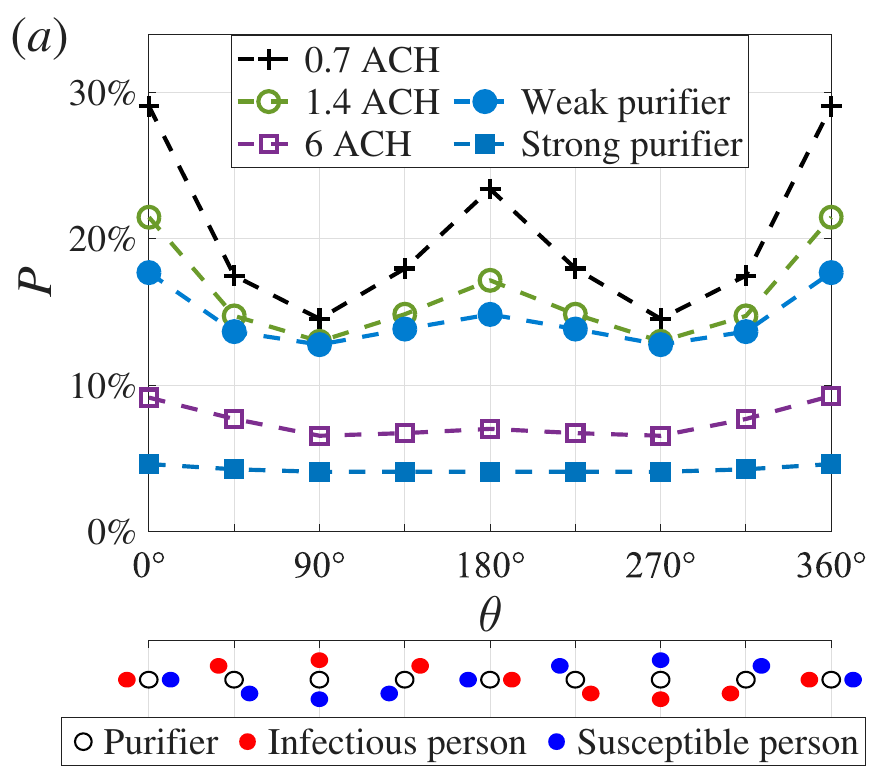}%{Fig6a.eps}
    \includegraphics[width=0.49\textwidth]{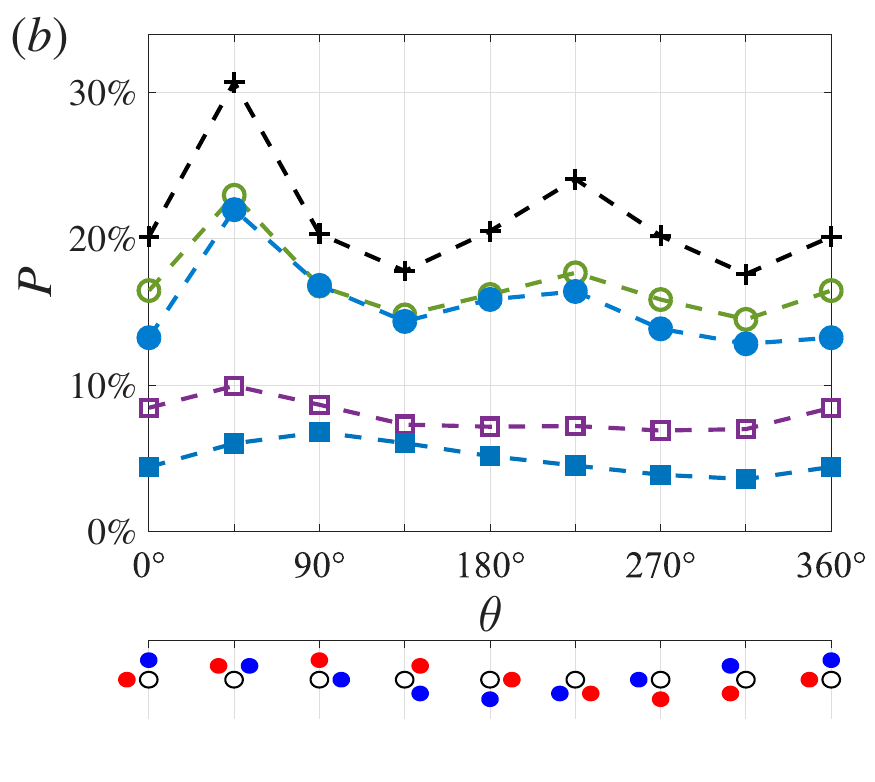}%{Fig6b.eps}
    \caption{The infection risk, $P$ \eqref{eq:InfRisk}, after 1 hour to the susceptible person at $\boldsymbol{x}_s$ \eqref{eq:SusLoc} with (\textit{a}) $\phi = 180^{\circ}$ and (\textit{b}) $\phi = 90^{\circ}$. Global ventilation cases are labelled with their ACH value (open symbols) and the equivalent local ventilation (purifier) is depicted by filled symbols of the same shape. Schematics in the lower axes show the locations of the infectious and susceptible people for each $\theta$.}
    \label{Fig6}
  \end{figure}

  To reflect a scenario in which the susceptible and infectious individuals are in close proximity, we consider $d = d_s = 1$ (note that $d = 1$ is the case with the greatest discrepancy between the local and global ventilation models in Figure~\ref{Fig4}). This is reflective of classrooms, lectures, and meetings, with 1 hour being a representative event duration. However, in reality, the strong purifier is potentially too large and noisy for use in such a setting.

  Two configurations of the susceptible and infectious people are considered: (i) they are opposite each other with the purifier in between them, $\phi = 180^{\circ}$ (Figure~\ref{Fig6}\textit{a}); and (ii) they are to one side of the purifier, $\phi = 90^{\circ}$ (Figure~\ref{Fig6}\textit{b}). Figure~\ref{Fig6} shows the infection risk to the susceptible person after 1 hour, with the scenario for each $\theta$ depicted on the lower axes. In Figure~\ref{Fig6}(\textit{a}), $P$ is symmetric around $\theta = 180^{\circ}$ because both the ADR equation and the airflow are symmetric in the line $y = L_y/2$. Moreover, the case $\phi = - 90^{\circ}$ can be deduced from Figure~\ref{Fig6}(\textit{b}) by reflection in this line ($\theta \rightarrow 360^{\circ} - \theta$).

  After 1 hour, the infection risk is below 40\% in all cases and the global ventilation model overestimates the infection risk. The weak purifier has had only a marginal effect when compared against the baseline example of 0.7 ACH, and shows close agreement with the corresponding global ventilation of 1.4 ACH. The infection risk is lower for the strong purifier than for the equivalent global ventilation of 6 ACH for all values of $\theta$. The greatest discrepancy is when $\theta = 0^{\circ}$, with the strong purifier predicting half the infection risk of the 6 ACH global ventilation (Figures~\ref{Fig6}\textit{a,b}).

  The streamlines of $\boldsymbol{v}$ again provide insight into these results. For the global ventilation models and the weak purifier, peaks occur when $y_0 = y_s$ due to the broadly unidirectional flow around the recirculating loop, with lower peaks when the aerosols travel further (e.g. $\theta = 180^{\circ}$ in Figure~\ref{Fig6}\textit{a}). For the strong purifier, the largest infection risk ($\theta = 90^{\circ}$ in Figure~\ref{Fig6}\textit{b}) corresponds to the susceptible person being directly downstream from the infectious person according to the significantly modified streamlines in this case (Figure~\ref{Fig2}\textit{b}).

  Under the WMR assumption, the infection risk is calculated based on the average aerosol concentration, $\bar{C}$ \eqref{eq:SSavg}. This approach predicts the following infection risks: 8.8\% for 0.7 ACH, 7.5\% for 1.4 ACH, and 3.6\% for 6 ACH. This is a significant underestimate compared to the corresponding values in Figure~\ref{Fig6} since the close proximity to the infectious source results in a concentration significantly greater than the room average at all times.

\section{Summary and conclusions} \label{Conclusions}

  The spatiotemporal modelling framework of \citet{Lau2022} for airborne disease transmission was modified to incorporate local ventilation effects motivated by air purifiers. For a weak purifier (CADR = 140 m$^3$h$^{-1}$) and a strong purifier (CADR = 1,000 m$^3$h$^{-1}$), the local ventilation model introduced here was compared against the global ventilation model \citep{Lau2022}, with 1.4 ACH and 6 ACH, respectively.

  For each purifier, the average aerosol concentration $\bar{C}$, after reaching a steady state, was compared for the local and global ventilation models, with good agreement in most cases (Figure~\ref{Fig3}). For the local ventilation model, $\bar{C}$ depends on the infectious source location and increasing the distance between the infectious person and the purifier reduces the average concentration. The lowest values of $\bar{C}$ were observed when the infectious source is inside the regions where the airflow streamlines are directed into the purifier inlet (Figure~\ref{Fig2}). When the infectious source is inside these regions, the global ventilation model significantly overestimates $\bar{C}$. When the infectious source is located outside of these regions, there is better agreement between the ventilation models, particularly for the weak purifier.

  When the infectious person is located inside or near to these regions, the global ventilation model overestimates the aerosol concentration everywhere in the room (Figure~\ref{Fig4}). However, when the infectious source is far from the purifier, there are some regions where the global ventilation model predicts larger concentrations, and other regions where the local ventilation model predicts larger concentrations (Figure~\ref{Fig5}). In these cases, the agreement between the local and global ventilation models is better for the weak purifier.

  The infection risk to susceptible people near the purifier after 1 hour was also considered (Figure~\ref{Fig6}). The weak purifier showed good agreement with the global ventilation model, whereas the strong purifier predicts a lower infection risk than the equivalent global ventilation (Figure~\ref{Fig6}). The largest discrepancies between the local and global ventilation models were observed when the susceptible person was located directly downstream from the infectious person according to the modified airflow streamlines (Figure~\ref{Fig2}). The global ventilation model overestimates infection risk in these cases.

  When modelling airborne transmission, Wells--Riley models \citep[e.g.][]{Miller2021} offer great computational speed but are very simple, whereas CFD models, particularly those that track individual particles, \citep[e.g.][]{Dbouk2021}, provide significant detail at greater computational expense. The modelling framework of \citet{Lau2022} offers a good compromise, including spatial variation at low computational cost. Here, we increase the complexity by developing a model able to explore the effects of local ventilation over hours, the time frame over which airborne transmission occurs, while still retaining relatively small computational times (around 5 minutes of simulation for 4-hour events).

  In future, these high computational speeds could allow a more thorough investigation, exploring factors such as purifier location, size, and strength. Also, the model could be further developed by incorporating a spatially varying eddy diffusion coefficient or an unsteady airflow. The local ventilation effects of air-conditioning, windows, doors, and other purifier designs could be explored through similar modifications to the modelling framework of \citet{Lau2022}.

\backsection[Acknowledgements]{The authors wish to acknowledge the contributions of Dr. Aaron English, Dr. Raquel Gonz{\'a}lez Fari{\~n}a and Dr. Attila Kov{\'a}cs. We are grateful to Si{\^ a}n Grant for generating Figure~\ref{Fig1}(\textit{a}).}

\backsection[Funding]{A.P. and K.K. gratefully acknowledge funding from a S{\^e}r Cymru `Tackling COVID-19' grant, awarded by the Welsh Government. I.M.G. is grateful to the Royal Society for funding through a University Research Fellowship.}

\backsection[Declaration of interests]{The authors report no conflict of interest.}

\backsection[Copyright]{For the purpose of Open Access, the authors will apply a CC BY copyright license to any Author Accepted Manuscript Version arising from this submission.}

\bibliographystyle{jfm}
\bibliography{Bibliography}

\end{document}